\documentclass[a4paper, times, 10pt, twocolumn, twoside,top=3.5cm,bottom=3.5cm,left=2.5cm,right=2cm]{article}

\usepackage{ISARC}
\usepackage{graphicx}
\usepackage{color}
\usepackage{CJK}
\usepackage{framed}
\usepackage{ulem}
\usepackage{amsmath}
\usepackage{caption}
\usepackage{subcaption}
\usepackage{bm}

\begin{document}

\renewcommand{\ISARCConference}{}

\definecolor{shadecolor}{rgb}{1,1,0}
\linespread{0.5}
\title{Automatic Crack Detection in Built Infrastructure Using Unmanned Aerial Vehicles}

\author{M.D. Phung, T.H. Dinh, V.T. Hoang and Q.P. Ha}
\affiliation{School of Electrical Mechanical and Mechatronic Systems, University of Technology Sydney, Australia}
\email{Manhduong.Phung@uts.edu.au, Quang.Ha@uts.edu.au}

\maketitle 
\thispagestyle{fancy} \pagestyle{fancy}

\begin {abstract}
This paper addresses the problem of crack detection which is essential for health monitoring of built infrastructure. Our approach includes two stages, data collection using unmanned aerial vehicles (UAVs) and crack detection using histogram analysis. For the data collection, a 3D model of the structure is first created by using laser scanners. Based on the model, geometric properties are extracted to generate way points necessary for navigating the UAV to take images of the structure. Then, our next step is to stick together those obtained images from the overlapped field of view. The resulting image is then clustered by histogram analysis and peak detection. Potential cracks are finally identified by using locally adaptive thresholds. The whole process is automatically carried out so that the inspection time is significantly improved while safety hazards can be minimised. A prototypical system has been developed for evaluation and experimental results are included.
\end{abstract}


\section{Introduction}
Crack detection is essential in health monitoring of infrastructure. Road sections containing a high density of cracks at the surface should be periodically maintained to ensure the safe operation of vehicles. Crack type, size and the level of severity need to be identified for this task. In concrete bridges, cracking caused by natural processes is inevitable and may result in malfunctioning of the entire bridge, or even collapse. It opens access for water, deicing salts and other corrosive chemicals to penetrate through the bridge deck and over time causes damages to internal bridge structures. Early identification of cracks is thus vitally important to maintain the service life of bridges.

To automatically detect cracks, both colour and geometry information need to be acquired with sufficient quality using ground mobile robots \cite{Kim2008, Prasanna2016}. However, complexity and roughness remain challenges for surface inspection using this approach. Recently, unmanned aerial vehicles (UAVs) have been developed as an alternative owing to their flexibility in the operation space and ability to carry specialised sensory equipment. In \cite{Eschmann2012}, a micro air vehicle system was employed to scan buildings using a high resolution camera. Images taken within a restricted location were then stitched with sufficient quality for crack and damage detection. An advanced UAV system was developed to evaluate the state of historical monuments \cite{Hallermann2013} whose captured images revealed after processing some damages in several monuments. In \cite{Metni2007}, a control system for navigating the UAV in unknown 3D environments was used to monitor and maintain bridges. UAVs were also used to inspect and monitor oil-gas pipelines, roads, power generation grids and other essential infrastructure \cite{Rathinam2008}.

For surface detection, computer-vision based techniques are widely used to detect cracks due to its robustness and cost efficiency \cite{Koch2015}. These techniques in general can be categorised into the wavelet transform, minimal path selection, machine learning, edge detection and intensity thresholding. For example, a separable 2D continuous wavelet transform is employed in \cite{Subirats2006}, using complex coefficient maps for crack segmentation. An improvement of the wavelet-based pavement distress detection can be achieved by combining the Wavelet-Radon transform and dynamic neural network thresholding \cite{Nejad2011}. These techniques however do not consider the geometric characteristics (orientation, continuity and connectedness) of the cracks and may wrongly detect the candidates with low continuity or high curvature. 

Owing to the ability to effectively identify high-level geometric information, the minimal path principle can be applied in surface crack detection. In \cite{Nguyen2011}, the free-form anisotropy is able to handle almost all crack characteristics in a segmentation step. An improvement of the minimal path technique is presented in \cite{Kaul2012}, having the capability to detect cracks without prior knowledge of endpoints. A fully unsupervised approach is proposed in \cite{Amhaz2016}, where a refined artefact filtering step is introduced to estimate the width of the crack. However, the main drawback of the minimal path approach is high computation time involved.

With the explosive development of image data, machine learning-based methods \cite{Oliveira2014, Shi2016} have been effectively used for surface crack detection. In \cite{Oliveira2014}, a multi-level pattern recognition system is developed to address image blocks containing cracks and then evaluate their geometry such as length and width. Random structured forests are used in \cite{Lim2013} and \cite{Shi2016} to extract information for crack descriptors from the image background. Nevertheless, these methods are greatly dependent on the training data, which is labour-intensive with manual labelling in the training and validation steps. 

For determining potential cracks from bridge decks, edge detection techniques is commonly used \cite{Adhikari2014, Li2014}. Four edge-detection algorithms are surveyed in \cite{Abdel2003}, where the Haar Wavelet method is identified as most reliable, compared to the gradient-based (Sobel and Canny) and frequency-based (fast Fourier Transformation) algorithms. These methods, however, only perform well under uniform illumination and low noise conditions.

In crack regions that are consistently darker than their surrounding  areas, intensity-thresholding methods \cite{Bi_Otsu} are more appropriate due to their compactness. A Bayesian classification technique together with a morphological opening operation and thresholding is applied to segment and classify defects from underground concrete pipes into various classes such as cracks, holes, laterals and joints \cite{Sinha2006}. A two-step method is proposed in \cite{Fujita2011} where a locally adaptive thresholding is used together with a median filter and a multi-scale line filter to emphasise the line structure and detect crack candidates. As thresholding is noise sensitive, it is often used with other techniques such as morphological or linear filtering to improve  robustness.

In this paper, we propose a system to inspect built infrastructure for automatic crack detection. A 3D model of the object is first created and its geometric features are extracted to generate a path for UAV navigation. While following the planned path, the UAV takes images of the suspected surfaces. Those images are then stitched and processed based on histogram analysis. For this task, we develop a peak detection algorithm for image clustering and a locally adjustable thresholding technique for crack detection. A number of experiments have been carried out and the detection results are promising to apply in real time applications.

\section{Data Collection Using UAV}

The goal of data collection is to acquire adequate geometric and surface information of the object to be used in post processing for detecting potential cracks, and also for navigation of the UAV itself \cite{Phung2016}. 

\subsection{Point cloud 3D modelling}
In this step, laser scanners are used to acquire range information from different positions of the structure to be inspected and represent them as point clouds. Those point clouds are then merged one by one in a process called registration to create a 3D model. In the registration, overlapping points corresponding to the same part of the structure appearing among the point clouds are first identified. Let $P_a$ and $P_b$ be the point clouds recorded at positions $\bm{a}$ and $\bm{b}$. The overlapping points are determined by:
\begin{equation}
\lVert (\bm{x}_i^a - \bm{a}) - (\bm{x}_j^b - b)\lVert < \tau,
\end{equation}
where $\bm{x}_i^a \in P_a$ and $\bm{x}_i^b \in P_b$ are overlapping points, considered in a close neighbourhood, and $\tau$ is a  pre-defined distance threshold. The alignment of point clouds is then obtained by applying an iterative closest point algorithm to find a transformation that minimises the distances between them.

\subsection{Geometric feature extraction}
Planar surfaces are often the main target to be inspected so they need to be extracted from the point cloud. Given a plane's equation $(ax+by+cz+d=0)$, then $M=[a,b,c,d]^T$ is the parameter vector to be identified. For this, a random sample consensus (RANSAC) algorithm is applied with some augmentations, including a noise filter to remove sparse outliers,  voxelisation to equalise the point density, and clustering to trim out the isolated groups before applying RANSAC. 

After detecting the surfaces, their boundaries are determined by using a convex hull algorithm. The remaining point cloud $P$ is then clustered into small groups as obstacles to be considered in path planning. Here, for a given positive constant $\epsilon > 0$, a cluster is defined as  a set of points: \begin{equation}
C = \{p_i \in P \, \vert \,~ \textnormal{min} \lVert p_i - p_j \lVert > \epsilon\},
\end{equation}
for any other point not belonging to the cluster, i.e., $p_j \notin C$.

\subsection{Path planning for colour image acquisition} 
Given surfaces to be inspected, a list of waypoints needs to be created to navigate the UAV. There are two types of waypoints, one corresponds to shooting points for taking images and the others serve as intermediate points for path following and avoiding obstacles. They are generated by first splitting the operating environment into voxels. A status of free or occupation is then defined for each voxel to indicate the existence or not of obstacles in that voxel. The shooting points are then computed based on intrinsic parameters such as camera focal length, surface area and minimum resolution. An A-star algorithm is finally applied to find the shortest path between shooting points. In each step, the cost to move from one voxel to another in the neighbourhood is computed as:
\begin{equation}
C(k,l,m)=a_1 k^2+ a_2 l^2+a_3 m^2, 
\end{equation}
where coordinates $k,l,m \in \{-1,0,1\}$  indicate the neighbour position, and the coefficients $a_1$, $a_2$ and $a_3$ assign a particular weight to each direction.
 
The generated waypoints are used as references for motion control of the UAV. Typically, the controller are built in the flight operating system so that the control task can be simplified. To collect images of sufficient quality, a gimbal is used to eliminate vibration and adds more degree of freedoms to the system to shoot images perpendicularly to the inspected surface. 

\begin{figure} [h!]
	\centering
	\includegraphics[width=0.5\textwidth]{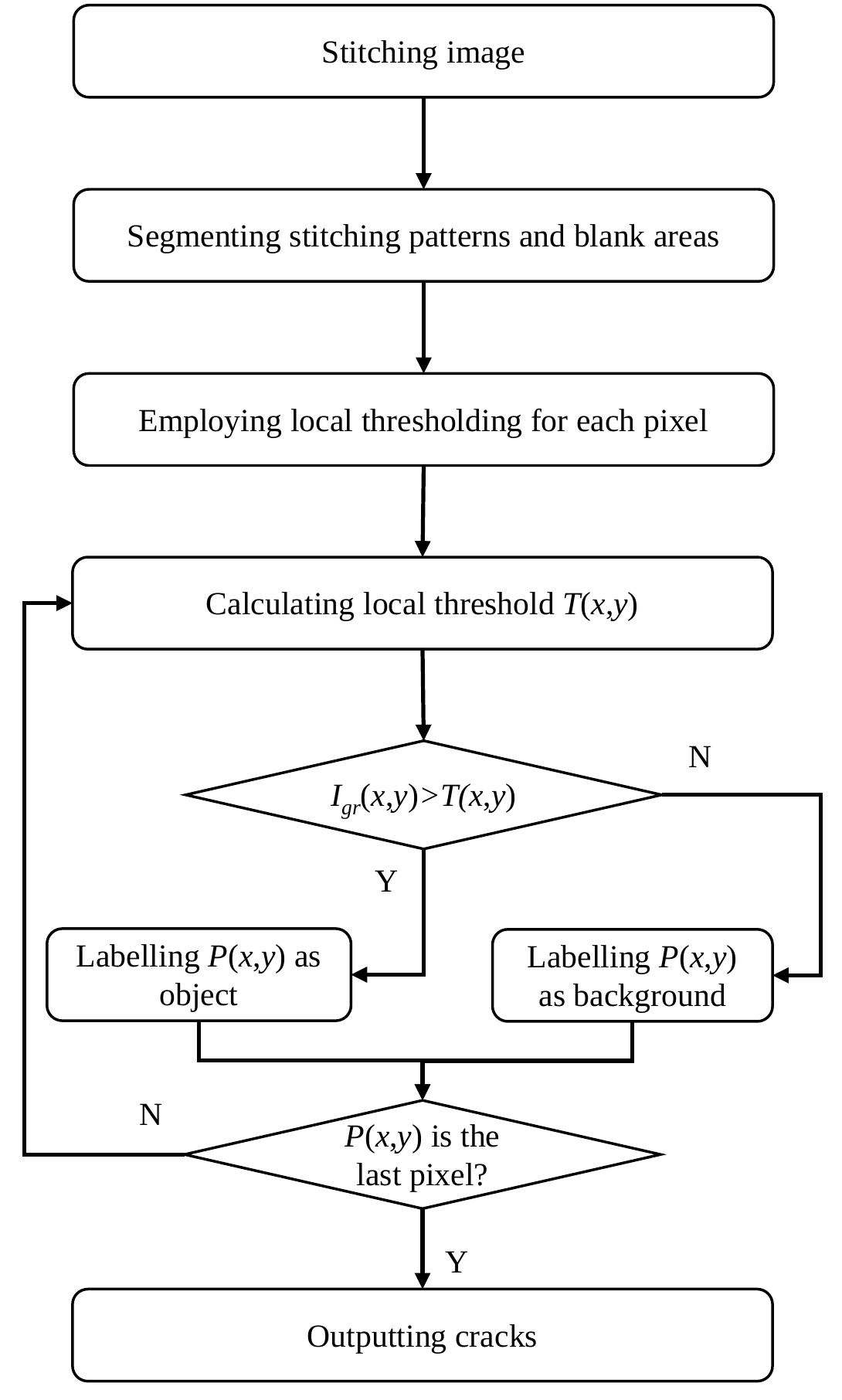}
	\caption{Crack detection flowchart.}
	\label{fig:Flowchart}
\end{figure}

\section{Thresholding-Based Crack Detection}
A crack detection algorithm is developed to process the collected data based on analysing their histogram information. Its flowchart including image stitching, pattern removal and crack detection steps is shown in Fig. \ref{fig:Flowchart}.

\subsection{Image stitching}
Each image taken by the UAV only covers a small area of the inspected surface. Stitching them to create a panoramic image is thus essential for crack detection. This stage requires a certain level of overlapping between consecutive images and corresponding features. An invariant feature based approach \cite{brown2007} is employed, consisting of scale-invariant feature transform (SIFT) feature extraction, homography computation and image matching verification. If the surface to be inspected is homogeneous, manual patterns can be added to enhance the stitching performance and removed afterwards. 

\subsection{Stitching pattern removal} 
Before processing the stitched image, it is important to remove the patterns used in stitching to enhance the detection accuracy. A stitched image typically consists of three elements, namely crack area, stitching patterns and blank area caused by alignment drifts. Since their luminance varies in accordance with the sunlight condition, our peak detection algorithm \cite{HiepICARCV16} is first employed to detect dominant peaks corresponding to these elements. The thresholds $t_1$ and $t_2$ to segment the surfaces to be inspected are calculated as the intensity average:
\begin{equation}
\label{eq:threshold}
\left\{\begin{array}{ll}
t_1=\dfrac{i_b+i_w}{2}\\
t_2=\dfrac{i_w+i_p}{2},
\end{array}\right.
\end{equation}
where $i_b$, $i_w$ and $i_p$ are the intensity value of peaks corresponding to the blank areas, the surface and stitching patterns, respectively. The stitching patterns are then identified based on the histogram as follows:
\begin{equation}
\label{eq:segmentation}
I_{gr}(x,y)=\beta \indent \text{if} \indent
\left [
\begin{array}{ll}
I_r(x,y)<t_1\\ 
I_r(x,y)>t_2,
\end{array}
\right.
\end{equation}
where $I_{gr}(x,y)$ and $I_r(x,y)$ are respectively the grey intensity and the red channel intensity of the stitched image at point $P(x,y)$, and $\beta$ is the intensity value chosen to distinguish the pattern with other parts of the image. As the intensity at crack structures is typically smaller than at non-defect areas of inspected surfaces, here $\beta$ is set to 255 for adequately filtering out those features without information loss.

\begin{figure} [h!]
	\centering
	\includegraphics[width=0.45\textwidth]{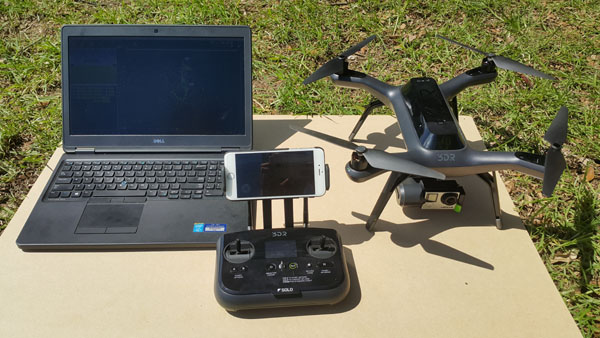}
	\caption{The 3DR Solo drone with remote controller and ground control station.}
	\label{fig:UAV}
\end{figure}

\subsection{Crack detection}
In an outdoor environment subject to varying lighting conditions, the global thresholding method \cite{HiepICARCV16} is extended to be able to extract all crack candidates with large variation of intensities. The approach proposed here, is first to take the advantage of the automatic peak detection for pre-processing the image to retain only the background and line-like objects, and then to apply locally-adjusted thresholds \cite{sauvola2000} to identify potential cracks.  Differing from the global approach, here the threshold is computed for each pixel based on the grey intensities of its neighbours. Let us consider $(x_i,y_i)$, $i = 1, 2, ..., N^2$, in the neighbourhood of pixel $P(x,y)$ determined by using an $N\times N$- window and $m(x,y)$ and $s(x,y)$ respectively as the mean and standard deviation in that window:

\begin{equation}\label{eq:adaptiveThresh}
m(x,y) = \frac{1}{N^2}\sum_{i=1}^{N^2}I_{gr}(x_i,y_i),
\end{equation} 

\begin{equation}\label{eq:adaptiveThresh}
s(x,y) = \sqrt{\frac{1}{N^2}\sum_{i=1}^{N^2}(m(x,y) - I_{gr}(x_i,y_i))^2}.
\end{equation} 
The threshold for pixel $P(x,y)$ is then computed as:
\begin{equation}\label{eq:adaptiveThresh}
T(x,y)=m(x,y)\left [1+k\left(\dfrac{s(x,y)}{R}-1\right)\right ],
\end{equation}  
where $R$ is the dynamic range of standard deviations and $k$ is a tunable parameter used to adjust the influence of standard deviation. Each pixel $P(x,y)$ is then evaluated against its threshold $T(x,y)$. A pixel $P(x,y)$ is considered as belonging to a crack if its grey intensity $I_{gr}(x,y)$ is higher than the computed local threshold, or lying in the background otherwise.

\begin{figure} [h!]
	\centering
	\includegraphics[width=0.45\textwidth]{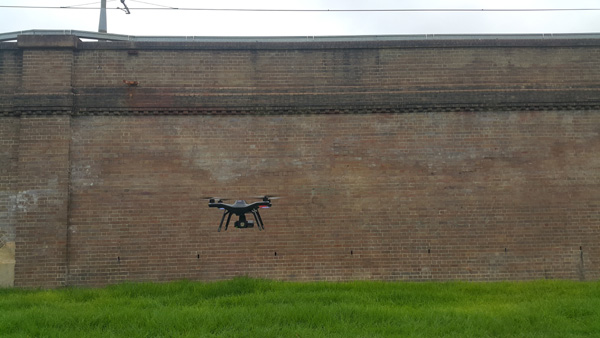}
	\caption{Collecting data of the bridge.}
	\label{fig:bridge_inspection}
\end{figure}

\begin{figure} [h!]
	\includegraphics[width=0.49\textwidth]{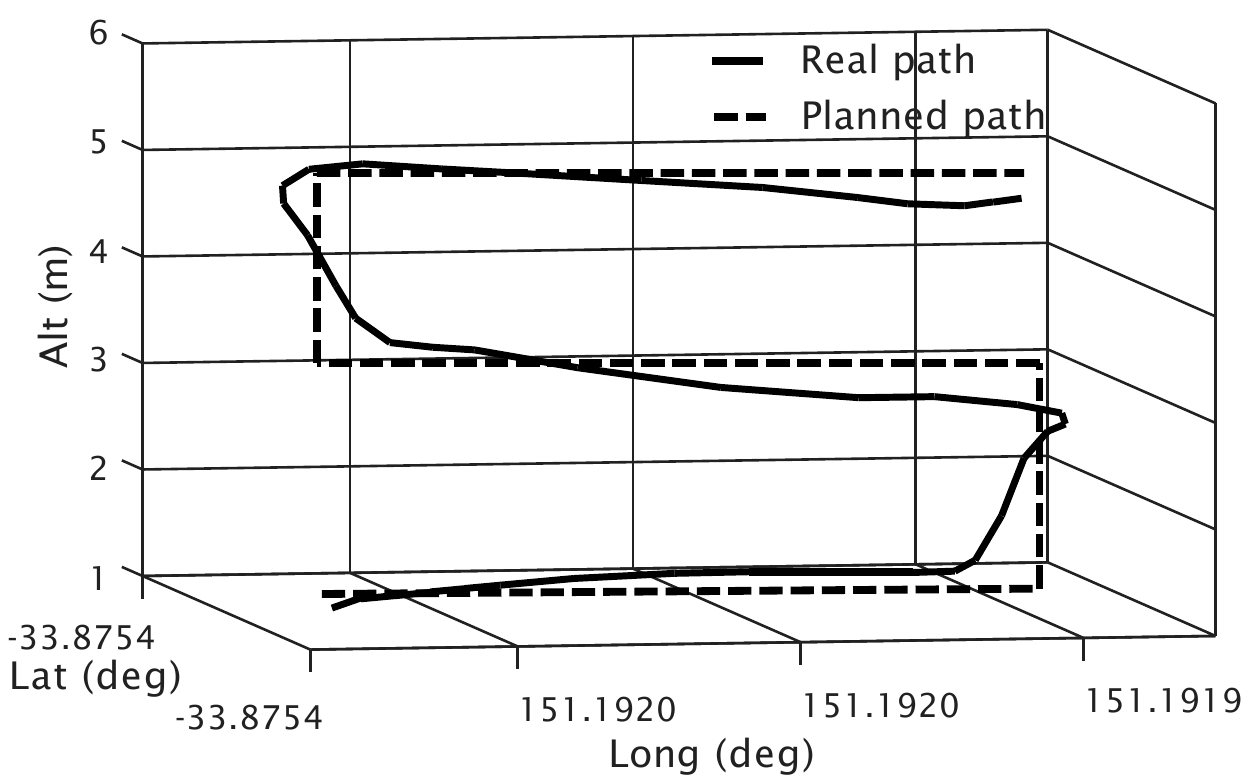}
	\caption{Flight path to collecting data of the bridge.}
	\label{fig:inpsection_path}
	
\end{figure}

\subsection{Experiments}
Experiments have been conducted to evaluate the proposed approach. The UAV used is a quadcopter, the 3DR Solo drone shown in Fig.\ref{fig:UAV}. It is equipped with three processors, two are Cortex M4 168 MHz running Pixhawk firmware for low-level control and the other is an ARM Cortex A9 running Arducopter flight operating system. The camera used is Hero 4 with the focal length of 34.4 mm and resolution of 12 megapixels, attached to a three-axis gimbal with one degree-of-freedom for controlling its yaw angle. The programming is carried out through the ground control station called Mission Planner and uploaded to the UAV. 

\subsubsection{Data collection results}
Figures \ref{fig:bridge_inspection} and \ref{fig:inpsection_path} show a practical path used to collect data of a bridge. There are mismatches between the planned and real paths which are inevitable due to localisation errors caused the built-in global positioning system. Nevertheless, in our experiments, the number of satellites detected ranging from 9 to 11 causing the error less than 1.5 m which is relatively small. For data collection, this error can be compensated by reducing the UAV's speed while increasing the number of shooting points. It is also noted that although the planned path shown in Fig. \ref{fig:inpsection_path} is rather ideal and does not consider the dynamic constraints of the UAV, which is beyond the scope of this study, the path tracking error remains however in an acceptable tolerance for the inspection purpose.       

\begin{figure} [h!]
	\setlength{\belowcaptionskip}{-20pt}
	\centering
	\includegraphics[width=0.45\textwidth]{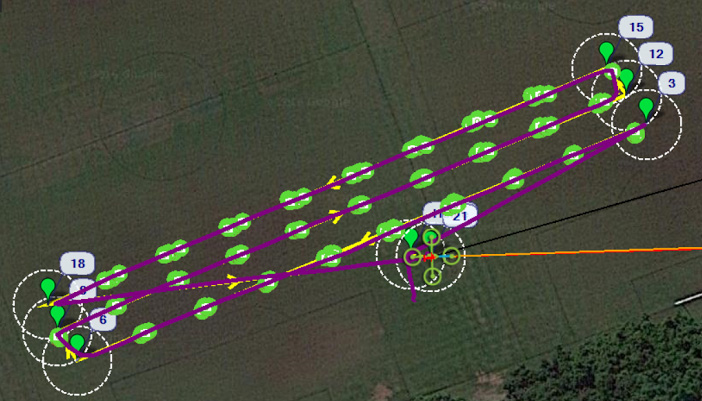}
	\caption{Flight path used to inspect the artificial wall.}
	\label{fig:inspection}
\end{figure} 

\subsubsection{Crack detection results}
\begin{figure} 
	\centering 
	\begin{subfigure}{0.45\textwidth}
		\centering
		\includegraphics[width=1\linewidth]{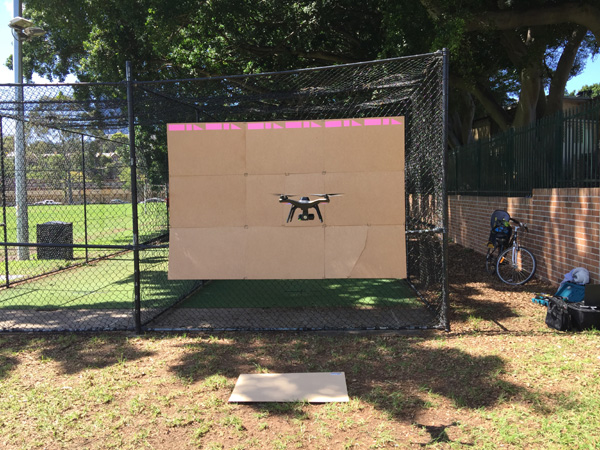}
		\caption{}
		\label{fig:cr1a}
	\end{subfigure}
	\begin{subfigure}{0.45\textwidth}
		\centering
		\includegraphics[width=1\linewidth]{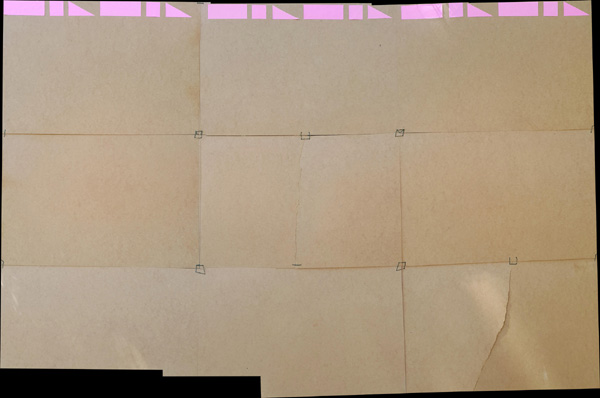}
		\caption{}
		\label{fig:cr1b}
	\end{subfigure}
	\begin{subfigure}{0.45\textwidth}
		\centering
		\includegraphics[width=1\linewidth]{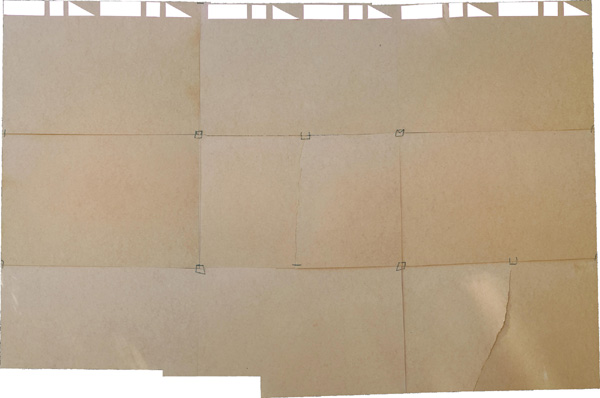}
		\caption{}
		\label{fig:cr1c}
	\end{subfigure}	
	
	\caption{Image stitching and segmentation:\\ (a) UAV taking photos of the artificial wall (b) Stitched image, (c) Segmented patterns and blank areas.}
	\label{fig:StiFil}
\end{figure}

To evaluate the crack detection algorithm, an wooden wall was set up from nine panels, each having the size of 600 mm $\times$ 900 mm $\times$ 3 mm. Those panels were hanged on a frame and joined together using twisted ties as shown in Fig. \ref{fig:cr1a}. Eighteen patterns were manually stuck on top of the panels for image stitching. The pink colour was chosen for the patterns due to its large difference from the colour of wooden panels. Two cracks were deliberately formed by a manual impact applied on two panels, one at the middle and one at the bottom right. The UAV was programmed to follow pre-defined waypoints to take images of the surface, as shown in Fig. \ref{fig:inspection}. The values of $R$ and $k$ in (\ref{eq:adaptiveThresh}) are chosen to be $128$ and $0.5$, respectively.

Figure \ref{fig:cr1b} shows the stitched image. It can be seen that there are almost no distortions compared to the original wall. However, missing pixels caused by alignment drifts still appear. This issue can be resolved by employing gain compensation and multi-band blending techniques. Figure \ref{fig:cr1c} shows the result of filtered stitching patterns. They are all well isolated from the stitched image demonstrating the effectiveness of the proposed automatic peak detection algorithm. 

\begin{figure}
	\centering
	\begin{subfigure}{0.45\textwidth}
		\centering
		\includegraphics[width=1\linewidth]{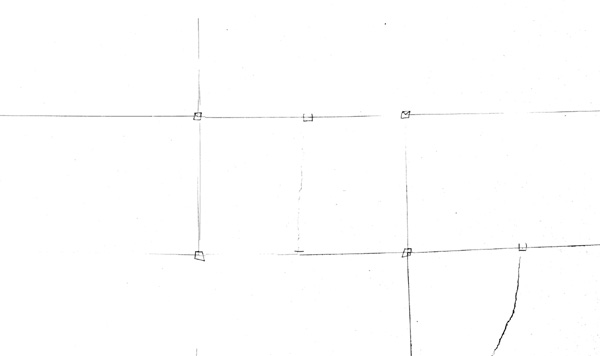}
		\caption{}
		\label{fig:cr2a}	
	\end{subfigure}
	
	\begin{subfigure}{0.45\textwidth}
		\centering
		\includegraphics[width=1\linewidth]{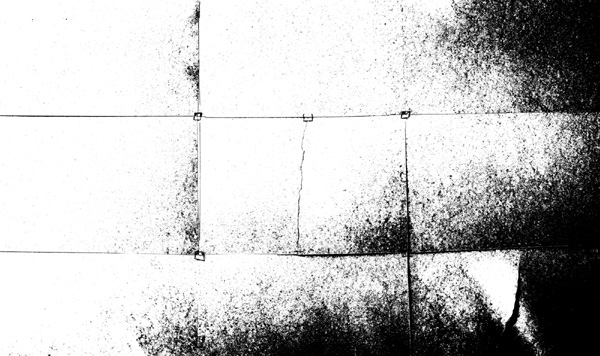}
		\caption{}
		\label{fig:cr2b}	
	\end{subfigure}

	\begin{subfigure}{0.45\textwidth}
		\centering
		\includegraphics[width=1\linewidth]{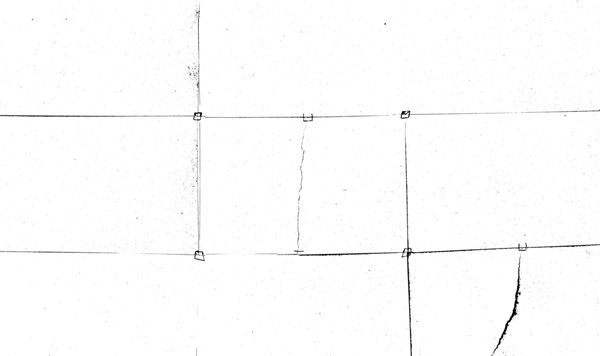}
		\caption{}
		\label{fig:cr2c}
	\end{subfigure}	
	
	\caption{Crack detection result: (a) Global thresholding at $\text{T}=125$, (b) Global thresholding at $\text{T}=155$, (c) Locally apaptive thresholding.}
	\label{fig:CroCra}
\end{figure}

Figure \ref{fig:cr2a} presents the crack detection results using the global thresholding method \cite{HiepICARCV16}. It can be seen that there are two crack candidates appearing along with horizontal and vertical lines of the tiled wooden panels. However, the middle crack is not well detected due to the variation in the grey intensity. If the threshold is increased to better capture the crack image, then parts of the wall that are exposed to the sunlight are not adequately filtered as shown in Fig. \ref{fig:cr2b}. This problem can be solved using the locally adjusted thresholding method as shown in Fig. \ref{fig:cr2c}.

\section{Conclusion}
In this work, we have developed an automatic crack detection system for infrastructure monitoring. By using UAVs, the system is capable of inspection of poorly accessible structures such as dams, culverts or bridges. A number of sensors have been integrated into the system allowing it to acquire geometric and colour information of the inspected surfaces. For data processing, we have developed computer vision based algorithms to create 3D models of the structure, extract features, plan navigation paths, stitch images and detect potential cracks.  A number of experiments have been conducted with all cracks detected in real time. Future work will focus on further improvements of the recognition algorithms to better identify crack properties such as the length, width and orientation.  

\section*{Ackowledgement}
The first author would like to acknowledge an Endeavours
Research Fellowship (ERF-PDR-142403-2015) provided
by the Australian Government. This work is supported
by a University of Technology Sydney Data Arena
Research Exhibit Grant 2016.

\bibliographystyle{plain}
\bibliography{ISARC}

\end{document}